%% file: coronaBF.tex
\documentclass[a4paper,9pt,conference]{IEEEtran}

\usepackage[utf8]{inputenc}
\usepackage[english]{babel}
\usepackage{algorithm}
\usepackage{algorithmicx}
\usepackage{amssymb}
\usepackage{amsmath}
\usepackage{graphicx}
\usepackage{textcomp}
\usepackage{url}
\usepackage{datetime}
\usepackage{verbatim}
\usepackage{enumitem}
\usepackage{amsfonts}
\usepackage{hhline}
\usepackage{adjustbox}
\usepackage{booktabs}
\usepackage{pifont}
\usepackage{multirow}
\usepackage{tikz}
\usetikzlibrary{shapes, positioning, arrows, decorations.markings, calc, patterns}
\usepackage{pgfplots}
\usepackage{boldline}
\usepackage{stmaryrd}
\usepackage{algpseudocode}
\usepackage{array}
\usepackage{forest}
\usepackage{tikz-qtree}
\usepackage{tkz-kiviat,numprint} 
\usetikzlibrary{arrows}
\usepackage[misc,geometry]{ifsym}

\newcommand{\m}{m}
\newcommand{\nkey}{k}
\newcommand{\A}{\mathcal{A}}
\newcommand{\B}{\mathcal{B}}

\newcommand{\K}{K}
\newcommand{\hmac}{h}
\newcommand{\bfA}{BF_{\A}}
\newcommand{\bfB}{BF_{\B}}

\newcommand{\nA}{n_{\A}}

\newcommand{\mail}[1]{\url{#1}}

\newtheorem{definition}{Definition}

\begin{document}

%\title{Identify Infection Chains with Bloom Filters Approach}
\title{Approach for GDPR Compliant Detection of COVID-19 Infection Chains}

\author{
\IEEEauthorblockN{Louis Tajan \qquad Dirk Westhoff}
\IEEEauthorblockA{%Institute ivESK\\
Hochschule Offenburg University of Applied Sciences\\
Offenburg, Germany\\
\mail{{louis.tajan, dirk.westhoff}@hs-offenburg.de}}

}

\maketitle

\begin{abstract}
While prospect of tracking mobile devices' users is widely discussed all over European countries to counteract COVID-19 propagation, we propose a Bloom filter based construction providing users' location privacy and preventing mass surveillance.
We apply a solution based on Bloom filters data structure that allows a 3\textsuperscript{rd} party, a government agency, to perform some privacy-preserving set relations on a mobile telco’s access logfile.
%By computing set relations, the government agency will identify infection chains from an initial and a final infected user and therefore be able to contact the intermediate possible infected users.
By computing set relations, the government agency, given the knowledge of two identified persons, has an instrument that provides a (possible) infection chain from the initial to the final infected user no matter at which location on a worldwide scale they are. The benefit of our approach is that intermediate possible infected users can be identified and subsequently contacted by the agency. 
With such approach, we state that solely identities of possible infected users will be revealed and location privacy of others will be preserved. To this extent, it meets General Data Protection Regulation (GDPR) requirements in this area.

%\begin{CCSXML}
%<ccs2012>
%<concept>
%<concept_id>10002978.10002991.10002995</concept_id>
%<concept_desc>Security and privacy~Privacy-preserving protocols</concept_desc>
%<concept_significance>500</concept_significance>
%</concept>
%<concept>
%<concept_id>10002978.10003014.10003017</concept_id>
%<concept_desc>Security and privacy~Mobile and wireless security</concept_desc>
%<concept_significance>500</concept_significance>
%</concept>
%<concept>
%<concept_id>10002978.10003029.10011150</concept_id>
%<concept_desc>Security and privacy~Privacy protections</concept_desc>
%<concept_significance>300</concept_significance>
%</concept>
%</ccs2012>
%\end{CCSXML}
%
%\ccsdesc[500]{Security and privacy~Privacy-preserving protocols}
%\ccsdesc[500]{Security and privacy~Mobile and wireless security}
%\ccsdesc[300]{Security and privacy~Privacy protections}

\end{abstract}

\input{introduction}
\input{scenario}
\input{approach}
\input{solution}
\input{conclusion}
%\begin{acks}
%	The discussions leading to this editorial were initiated during
%	Dagstuhl Seminar 15102 on
%	\emph{Secure Routing for Future Communication Networks},
%	and we thank all participants for their contributions.
%\end{acks}

\bibliographystyle{unsrt}
\bibliography{references}

\end{document}

%% file: introduction.tex
\section{Introduction}
Cases of COVID-19 disease have been reported in more than 190 countries and its spreading has been characterized as pandemic by the \textit{World Health Organization} on 11.03.2020.
One of its multiple side effects consists of European democracies being challenged. Indeed, several countries are collecting location-based data from their own citizens. The \textit{state of emergency} for health reasons has been established in countries as Spain, Portugal, France or Switzerland. Such a specific situation empowers a government to perform actions that would normally not be allowed to undertake.
For instance, in Milano, Italy, mobile network operators are providing information on users' traffic to public authorities. In Germany, issues regarding how and for which usage to process the location-based information are ones of the most discussed. Indeed, efforts in Germany are twofold regarding digital support to detect infection chains.
First, with an app \textit{Corona-Warn-App} deployed and downloaded more than 15 millions times in Germany (population of approx. 83 millions). It consists of using a tracking app with Bluetooth in which a smartphone of an infected user is subsequently informing all devices which have been in proximity (within the beaconing received range at some point in time in the past). Such an approach is very vulnerable due to the requirement of continuously activated Bluetooth. The recently published families of BIAS \cite{antonioli20bias} or BlueBorne \cite{DBLP:conf/fmec/AlmianiRYKMAA19} attacks have shown that mobile devices with activated Bluetooth can easily be remotely executed, e.g. \textit{CVE-2017-078}1, \textit{CVE-2017-0782} or \textit{CVE-2017-14315} and are classified as a severe risk.
Moreover, it has been pointed out that the harvesting of contacts via Bluetooth with a tracking app is only properly working
in case the app is activated continuously in the foreground, and, moreover, that at least $60\%$ of the smartphone users
need to download and continuously use it to indeed have an impact with respect to the identification of infection chains.

Second, telco operators would provide access logfiles of mobile network base stations to RKI (Robert-Koch-Institute) to support inferring infection chains.

On the contrary, the Netherlands' government decided to not approve a general confinement, for the reason of being incompatible with individual freedom.

For these reasons, in the work at hand we attempt to propose a construction which combine the efficiency to help the public authorities to contain the virus spreading with the possibility to provide privacy with respect to the citizens. Therefore, we concentrate on providing a privacy-preserving solution for the 2nd effort currently done within Germany.
Our proposed solution makes use of our previous works \cite{DBLP:journals/iacr/TajanWA19,DBLP:conf/icete/TajanWA20} which allows a non-trusted third party to privately compute operations and relations on sets using Bloom filters data structure. Such data structure allows one to represent a large set of elements in a simple tabular of bits which could provides obfuscation and privacy on the set.

We recall that GDPR's two main objectives are to firstly enhance the personal data protection by processing them and to secondly empower the companies in charge of this processing procedure. Even if this regulation does not apply on fields as public health or national security \cite{HoofSB20}, weaving the proposed Bloom filter based private protocols into infection chains investigation would limit government agencies to solely identify users with high probability of being infected instead of a massive data analysis of all mobile users.

\subsection{Related Work}
Several approaches from related work allow one to perform computations on pseudonymized, obfuscated or even encrypted data without the need to discern them. We could list homomorphic encryption \cite{DBLP:conf/eurocrypt/FreedmanNP04, DBLP:conf/ccs/Kerschbaum12} or multi-party computation \cite{DBLP:conf/crypto/KissnerS05, DBLP:conf/csreaSAM/LaiYCCH06} which represent the mainly investigated techniques. In \cite{DBLP:conf/cecc/TajanW19}, we applied our Bloom filter based construction to several use cases of post-mortem mobile device tracking.
In our former work \cite{DBLP:conf/icete/TajanWA20}, we have shown that this alternative approach based on Bloom filter could be used to secure data while preserving the ability of performing relevant tests or computations on the private data. Bloom filters have been used in many different scenarios as presented in \cite{DBLP:journals/comsur/LuoGMRL19}. For instance, Kerschbaum directly encrypts the Bloom filter with homomorphic encryption \cite{DBLP:conf/ccs/Kerschbaum12}. In \cite{DBLP:conf/coinco/MauryaS17} authors applied the Bloom filter to key exchange mechanisms in wireless sensor network (WSN) environment while in \cite{DBLP:conf/icoin/TalpurNSSFK17}, authors optimize the sensor nodes broadcasting with the use of Bloom filters.

Regarding the investigation of privacy-preserving location tracing solutions in the environment of COVID-19 spreading, we could mention the work of PEPP-PT consortium \cite{PEPP:2020}. This European team provide standards, technology, and services to countries and developers with the objective to help stopping the COVID-19 spreading.

%% file: scenario.tex
\section{Scenario}
%We have two users $A$ and $B$ who are infected by the COVID-19 virus.
A government agency, which role is to reduce the spreading of the SARS-CoV-2 virus in its country, knows different pairs of infected persons $(A,B)$. Its objective here, is to identify all the possible paths which relies user $A$ to user $B$ and considers the case where infection of user $B$ is a consequence of user $A$'s infection. By retrieving all possible paths (surely it could also turn out that no path exists and the infection of users $A$ and $B$ was unrelated), the agency could identify all the users within this path that may be also infected by the virus and try to contact them. Indeed, different mobile device's users close to the same mobile base station at the same time could potentially spread the virus in case of one being infected.
To do so, the agency is analyzing connection data provided by a telco company. The connection logs are collected on the base stations which are providing network access to the users' mobile devices.

\subsection{Parties Involved.}
Four parties are involved in the scenario:
\begin{description}
\item[Users:] could be infected by the SARS-CoV-2 virus. They are connecting to the base stations to access the mobile network.
\item[Telco company:] provides network to the users via several base stations. It also provides log data from the network connections to government agencies.
\item[Base stations:] are distributed over several countries, provide network to the users' mobile devices and collect connection data.
\item[Government agency:] aims to identify "infection chains" in order to contact the possible infected users and counteract the COVID-19 disease pandemic.
\end{description}

\subsection{Collecting Connection Data}
For each base station $j$, the telco company firstly generates and initializes a fresh Bloom filter $BF_j$ represented by a tabular of bits, all set to 0. Any time a user is connecting to the mobile network using base station $j$, the following connection information is aggregated and added to $BF_j$:
$$(id_i, t^1_i, t^2_i)$$ with $id_i$ the user's credentials and $t^1_i$ and $t^2_i$ respectively the starting and ending times of its connection to the access point. Such connection data should be considered as sensitive regarding the location privacy of the users.
As it will be presented, we consider a Bloom filter-based approach which brings privacy to the stored data. Indeed, on the one hand the base stations are using usernames to characterize the users and on the other hand only the telco company could generate and access the connection information from the base stations.

\subsection{Proximity Chain - Infection Chain}
\textit{As notation rule, we use $\langle \rangle$ to express proximity chains and $[ ]$ for infection chains.}\\
A proximity chain consists of a list of users where two successive ones have been at the same location at the same time. To establish a proximity chain, these times of contact should be ordered. In other words, in the proximity chain $\langle A, D, F, E, B\rangle$, the time at which users $A$ and $D$ have been at the same location should precede the one for users $D$ and $F$ (i.e. $[t^1_A;t^2_A] \cap [t^1_D;t^2_D] < [t^1_D;t^2_D] \cap [t^1_F;t^2_F]$).

In addition to be defined as a proximity chain, the list could also represent an infection chain. In this case, all the users composing the chain should have a probability of being infected $Pr(X_i)$ greater than a certain threshold $Tr$.
More concretely, an infection chain $[A, X_1, \dots, X_n, B]$ is a proximity chain for which it holds that:
$\forall \ X_i: Pr(X_i)>Tr$, otherwise it is solely a proximity chain.
Therefore, an infection chain $[A, X_1, \dots, X_n, B]$ represents how the SARS-CoV-2 virus may have spread from an initially infected user $A$ to a consecutive infected user $B$. 

It may happen that one or several subsets of a proximity chain $\langle A, X_1, \dots, X_n, B\rangle$ are considered as infection chains, e.g.\\ $[A, X_1, \dots, X_i]$ and/or $[X_j, \dots, B]$.

\subsection{Adversary Model:}
We consider the government agency as the principal threat for the application's users. As we stated previously, even if GDPR does not apply on public health security matters, we aim to apply limitations on government agencies. In such a way, we would like that the agencies could only identify users with high probability of being infected instead of having a massive data analysis of all mobile users. As we will present in the following sections, having the telco company colluding with the government would allow the agency to access personal data of all users and therefore we do not consider such assumption.

Even if we do not get any collision, we could also precise that users are not trusting the telco company. Indeed, they seek to limit the mobile devices to collect personal data as much as possible.

We also consider that users do not trust any approaches that require to maintain Bluetooth continuously on since multiple types of attacks could occur as by example remote code executions from \textit{Bleedingbit} vulnerabilities \cite{bleeding}.
%Sensor nodes are not equipped with tamper-proof hardware, and the adversary is capable of compromising and fully controlling arbitrary number of sensor nodes, therefore we assume all the nodes, namely sensor or intermediate, as \textit{malicious} parties. On the contrary, we assume that $\bs$ is immune to all physical attacks and it is trustworthy. We thus consider it as an \textit{honest-but-curious} party. Moreover, the adversary can eavesdrop and alter any messages from honest nodes. In a word, we consider an adversary granted with a full-scale attack capacity against the sensor network except $\bs$.

%% file: approach.tex
\section{Bloom Filters-based Approach}
As recently proposed in \cite{DBLP:conf/icete/TajanWA20}, the Bloom filter data construction could allow to privately represent sets of elements and at the same time enable performance-saving computation on them. Exactly due to this performance-saving privacy extension, we argue that our approach also suits for such massive data sets like mobile access logfiles. At next, we give a background on Bloom filters and the relevant set relation and recall the basic protocol's functions.

\subsection{Bloom Filters}
A Bloom filter is a data structure introduced by Burton Howard Bloom in 1970 \cite{Bloom:1970:STH:362686.362692}. It is used to represent a set of elements. With a Bloom filter representing a certain set, one can verify whether an element is a member of this set. Such a data structure consists of a tabular of $\m$ bits which is associated to $\nkey$ public hash functions. At first, all the $\m$ bits are initialized to 0.
%Moreover two functions namely add() and test() are available.
To add an element to the Bloom filter, one has to compute the hashes of this element with each of respective $\nkey$ hash functions. Then, set the bit to 1 for each position corresponding to a hash value. To test whether one element is included in the Bloom filter, one has, similarly, to compute the respective hash values of this element and verify if the respective bits are set to 1. If at least one of these bits is set to 0, then we know for sure that the tested element is not a member of the set represented by the Bloom filter (i.e. no false negative could append when testing an element). On the contrary, with some (minor) probability, the testing function could retrieve a false positive. Indeed, even if all the bits that have been verified are set to 1, the tested element may not be part of the set represented by the Bloom filter.

\subsection{Set Relations}
Multiple types of operations could be performed on sets.
%In this work, we aim to test set relations and some of them could be reduced to compute the cardinality of some operations. For instance, being able to compute the cardinality of the intersection of two sets indicates whether they have elements in common or if one set is included in the other one.
For privacy concerns it could be of interest to solely reveal the cardinality of the resulting set instead of its content. Therefore, we propose a solution on adapted Bloom filters (see \ref{subsec:functions}) to use one kind of set relations namely the inclusiveness defined as follows:
\begin{definition}[Inclusiveness]
Let $\A$ and $\B$ be finite sets. We consider $\A$ included in $\B$, i.e. $\A\subset \B$, iff all elements from $\A$ are included in $\B$ : $\forall a \in \A : a\in \B$.
\end{definition}
%\vspace{-3mm}
%\begin{definition}[Disjointness]
%Let $\A$ and $\B$ be finite sets. We consider $\A$ and $\B$ as disjoint iff none of the elements from $\A$ are included in $\B$. In other words, $\A \cap \B = \emptyset$ : $\forall a \in \A : a \notin \B$.
%\end{definition}

\subsection{Private Protocols}\label{subsec:functions}
To guarantee full privacy of the sets' content along with their cardinality, we proposed in \cite{DBLP:conf/icete/TajanWA20} to modify the Bloom filters approach in two aspects. Firstly, instead of using $\nkey$ public hash functions, we are using a unique HMAC function with $\nkey$ secret keys. Secondly, the exact value of $\nkey$ is kept secret and is privately and randomly generated within two publicly known boundaries. We specify the functions regarding the initialization phase and the inclusiveness protocol.
\subsubsection{Initialization.}
\begin{description}
\item[$\mathbf{\hmac, \nkey, \m, \K  \leftarrow}$Setup:]
The telco company should first choose and generate the Bloom filter parameters: the dimension $\m$, the HMAC function $\hmac$, the amount of keys $\nkey$ and the set of keys $\K =\{\kappa_1, \dots , \kappa_{\nkey}\}$.
\item[$\mathbf{\bfA \leftarrow}$Create($\hmac, \m, \K, \mathbf{\A}$):]
Generates the Bloom filter of the data set $\A=\{a_1, \dots, a_{\nA}\}$.
\end{description}

\subsubsection{Inclusiveness Protocol.}\label{subsec:inclusiveness}
This operator allows to verify if one set is included in another. It performs directly on the Bloom filters of the respective sets.
This operator is defined as:
\begin{description}
\item[$\mathbf{BF_{\A \subseteq \B} \leftarrow INC(\bfA, \bfB)}$:]
For each index, we set 0 if at the same index we have 1 for $\bfA$ and 0 for $\bfB$ and we set 1 otherwise.
\end{description} 
This operator is equivalent to the bitwise binary operator combination:
\begin{equation}
INC(BF_\A, BF_\B) \equiv \neg (BF_\A) \ OR \ \ BF_\B 
\end{equation}

Then we express the number of bits set to 1 in the resulting Bloom filter. If it is equal to $\m$, we can conclude that $\A\subseteq \B$ if no false positive occurred. Otherwise we get $\A\nsubseteq \B$ with certainty.

%\subsubsection{Disjointness Protocol.}\label{subsec:disjointness}~\\
%This set relation allows to verify that no elements from one set are included in another set. In other words, this allows to claim that two sets are disjoint. This test function is not trivial, indeed, if we use Bloom filters it is not sufficient to highlight the cases where a bit 1 has been inserted at the same index for the two respective Bloom filters.
%The operator is defined as:
%\begin{description}
%\item[$\mathbf{BF_{\A \cap \B=\emptyset} \leftarrow DIS(\bfA, \bfB)}$:]
%For each index, we set 1 if at the same index we have 1 for both $\bfA$ and $\bfB$ and we set 0 otherwise.
%\end{description}
%This operator is equivalent to the bitwise logical-and operator:	
%\begin{equation}
%DIS(BF_\A, BF_\B) \equiv BF_\A \ AND \ BF_\B.
%\end{equation}
%Then we express the number of bits set to 1 in the resulting Bloom filter $\mathbf{BF_{\A \cap \B=\emptyset}}$ and if it is less than the lower bound of $\nkey$ then $\A$ and $\B$ are distinct and $\A$ and $\B$ have at least one element in common otherwise.
%Indeed, for each element which is included in both sets, we get $\nkey$ times a bit set to 1 in the resulting Bloom filter. However we could still get such a bit set to 1 due to a bit set to 1 in $\bfA$ and $\bfB$ stemming from different elements originally added to the Bloom filters. We call such a case a false positive for the disjointness relation.

For an evaluation of the correctness and the security of this protocol, we refer the readers to \cite{DBLP:journals/iacr/TajanWA19}. It is shown that a proper selection of parameters $\m$ and $\nkey$ considering the number of elements to be inserted, guarantees the limitation of overlapping bits in the resulting Bloom filter and enables the 3\textsuperscript{rd} party to correctly conclude on the inclusiveness property of the two sets. Indeed, a too large amount of overlapping bits in the resulting Bloom filter would lead to a case of false negative.

%Also, in \cite{DBLP:journals/iacr/TajanWA19} the authors aim to have a sufficient amount overlapping bits in the objective of obfuscating the cardinality of the sets while observing their respective Bloom filter representations.

%% file: solution.tex
\section{Proposed Solution}

%\subsection{Big Picture}
%\begin{enumerate}
%\item identify all chains $\{A, X_1, \dots, X_m, B\}$ from given users $A$ and $B$.
%\item use Gossip formula to model probability of infection when two nodes $X_i$ and $X_{i+1}$ are in proximity.
%\item apply $Pr(E^1_p \ AND \dots AND \ E^n_p) = e^n$ (from \cite{DBLP:journals/sigmobile/LamparterPW03}) to estimate if user $A$ has indeed been infected by user $B$ (here $E^1_p$ can be modeled with the Gossip probability of step 2).
%\item in case a threshold $Tr < Pr(E^1_p \ AND \dots AND \ E^n_p)$, reveal the chain $\{A, X_1, \dots, X_m, B\}$ and contact each intermediate user.
%\end{enumerate}

From any two given infected users $A$ and $B$, the government agency first aims to identify all the proximity chains $\langle A, id_{X_1}, \dots, id_{X_n}, B\rangle$. In our protocol, we recall that the telco company provides all the relevant Bloom filters to the government agency. We propose to dissociate three cases:
\begin{itemize}
\item \textbf{CASE 1}: the smallest possible proximity chain $\langle A, B\rangle$:\\
  there is a base station $BS_j$ and a Bloom filter $BF_{A,B}$ for set $\{A,B\}$ and $INC(BF_j, BF_{A,B})=true$.\\
  Since both users $A$ and $B$ are indeed infected, the proximity chain $\langle A ,B\rangle$ is also an infection chain $[A, B]$.
    
\item \textbf{CASE 2}: a proximity chain with one intermediate user $X$ $\langle A, id_{X}, B\rangle$:\\
 there is a base station $BS_{j_1}$ and a Bloom filter $BF_{A,id_{X}}$ for set $\{A, id_{X}\}$ where $INC(BF_{j_1}, BF_{A,id_{X}})=true$  
 and in addition, there is a base station $BS_{j_2}$ and a Bloom filter $BF_{id_{X}, B}$ for set $\{id_{X}, B\}$ and $INC(BF_{j_2}, BF_{id_{X},B})=true$.
We remark here that we know users $A$ and $B$ but we do not know user $X$ nor his access credential $id_{X}$, so the government agency has to search in all base stations for all $X_j$ for which the above two inclusiveness tests INC hold.\\
If $Pr(id_X)>Tr$ we can denote $[A, id_X, B]$.
    
\item \textbf{CASE 3}: the general case $\langle A, id_{X_1}, \dots, id_{X_n}, B\rangle$:\\
 we have $INC(BF_{j_1}, BF_{A,id_{X_1}})=true$ $\land$ $\dots$ $\land$\\ $INC(BF_{j_n}, BF_{id_{X_n},B})=true$.
\end{itemize}

Our solution consists of having the government agency building a data tree structure representing all the proximity chains starting from user $A$. From this tree, the agency could easily identify the proximity chains from user $A$ to user $B$. For the next step of the protocol, the government agency has to evaluate the chain to determine its plausibility to actually be an infection chain. We give the outlines of this step but not its evaluation function that we save for the epidemiologists. 

We emphasize that at this point, the proximity or infection chains will only reveal usernames of users $X_1, \dots, X_n$ and not their real identities. At the very end of the protocol, the government agency will request from the telco company the identities of the intermediate infected users.

\subsection{Generating the Proximity Tree}
To obtain a proximity tree, the government agency starts by creating an empty tree $T$ with user $A$ as root. Then, it processes the recursive algorithm $prox\_tree(A, A, B, t^\prime)$ presented in Algorithm \ref{algo:F} with $t^\prime$ the time from when user $A$ could have started the infection process. The recursive algorithm does as follow: first, it generates the list $BS_N$ of base stations that the current node $N$ has been connected to at a time later than $t$.
To test if a user $N$ has been connected to a base station $j$ (i.e. test if $(id_N, t^1_j, t^2_j) \in BF_j$), the government agency receives from the telco company all the Bloom filters composed of each of the 3-tuples $(id_N, t^1_j, t^2_j)$. Then, the government agency performs the inclusiveness testing between the received Bloom filters and $BF_i$, the Bloom filter corresponding to the connections logfile from $BS_j$ as: $INC(BF_{N,j}, BF_i)$. The next step of the algorithm consists of identifying all the users that visited the base stations from set $BS_N$ at the same moment than user $N$. As before, the telco company generates Bloom filters with the 3-tuples $(id_l, t^1_{l}, t^2_l)$ for all users $l$ and all time ranges $[t^1_{l}; t^2_l]$ that overlap the connection time of user $N$. To determine which users should be listed, the government agency performs the inclusiveness operator between these Bloom filters and $BF_N$ the one composed by the elements from $BS_N$. Finally, for every identified users, they are added to the proximity tree $T$ as a leaf of current node $N$ and Algorithm \ref{algo:F} is then recursively processed on the leaves.

An additional aspect to take into account while recursively processing the algorithm is to consider the upper nodes of the current node in the proximity tree. Indeed, we would like to avoid creating some loops in the tree which are irrelevant when dealing with infection problems; if user $A$ infected user $C$, it makes no sense to consider user $C$ infecting user $A$ in short period of time.
The algorithm should then exclude all the users which are already inserted as upper nodes in the tree. Regarding the tree construction, if we consider that user $C$ has been in proximity of user $A$ and $id_C$ is added as a leaf of root $A$, user $A$ should not be considered anymore as potential leaf of node $id_C$ and so on.

\begin{algorithm}
\caption{$prox\_tree(N, A, B, t)$}\label{algo:F}
\begin{algorithmic}[1]
\Require a node N from a tree T, users $A$ and $B$, a time $t$
\Ensure a tree T
\If {$N=B$} 
        \State \textbf{break}
\EndIf  
\ForAll{$BF_j$}
	\State{ \ForAll{$t^1_j,t^2_j>t$}
		\State{ \If{$(id_N, t^1_j,t^2_j) \in BF_j$}
			\State {$BS_N.add((BS_j,t^1_j,t^2_j))$}
		\EndIf}
	\EndFor}
\EndFor

\ForAll{$(BS_{Nk},t^1_{Nk},t^2_{Nk}) \in BS_N$}
	\State{ \ForAll{$id_l$}
		\State{ \ForAll{$(t^1_l, t^2_l)$ $\mid$
	($t^1_{l}\leqslant t^1_{Nk}$ $\land$ $t^2_{l}\geqslant t^1_{Nk}$) $\lor$ ($t^2_{l}\geqslant t^2_{Nk}$ $\land$ $t^1_{l}\leqslant t^2_{Nk}$) $\lor$ ($t^1_{l}\leqslant t^2_{Nk}$ $\land$ $t^2_{l}\geqslant t^1_{Nk}$)}
			\State{ \If{$(id_l, t^1_{l}, t^2_l) \in BF_{Nk}$}
	  			\State $createLeaf(id_l)$ \label{line20:algo}
	  			\State $prox\_tree(id_l, A, B, max(t^1_{Nk}, t^1_l))$
			\EndIf}
		\EndFor}
	\EndFor}
\EndFor
\If {$N.leaf= \emptyset$}
		\State \textbf{break}
\EndIf
\end{algorithmic}
\end{algorithm}

%\begin{algorithm}
%\caption{Calculate $y = x^n$}
%\begin{algorithmic}
%\STATE INPUT: $n \geq 0 \vee x \neq 0$
%\ENSURE $y = x^n$
%\STATE $y \leftarrow 1$
%\IF{$n < 0$}
%\STATE $X \leftarrow 1 / x$
%\STATE $N \leftarrow -n$
%\ELSE
%\STATE $X \leftarrow x$
%\STATE $N \leftarrow n$
%\ENDIF
%\WHILE{$N \neq 0$}
%\IF{$N$ is even}
%\STATE $X \leftarrow X \times X$
%\STATE $N \leftarrow N / 2$
%\ELSE[$N$ is odd]
%\STATE $y \leftarrow y \times X$
%\STATE $N \leftarrow N - 1$
%\ENDIF
%\ENDWHILE
%\end{algorithmic}
%\end{algorithm}

%The set $BS_N$ represents the list of all the base stations that user $N$ visited at a time later than $t$.

%In Figure \ref{fig:tree} we give an example of a proximity tree obtained by the government agency after computing the recursive function for users $A$ and $B$. As a result from this case, if proximity chains $\langle A, E, F, B \rangle$ and $\langle A, D, B \rangle$ are characterized as a infection chain then users $D, E$ and $F$ might also be infected.

\begin{figure}[h]
\begin{center}
\includegraphics[width=0.47\textwidth]{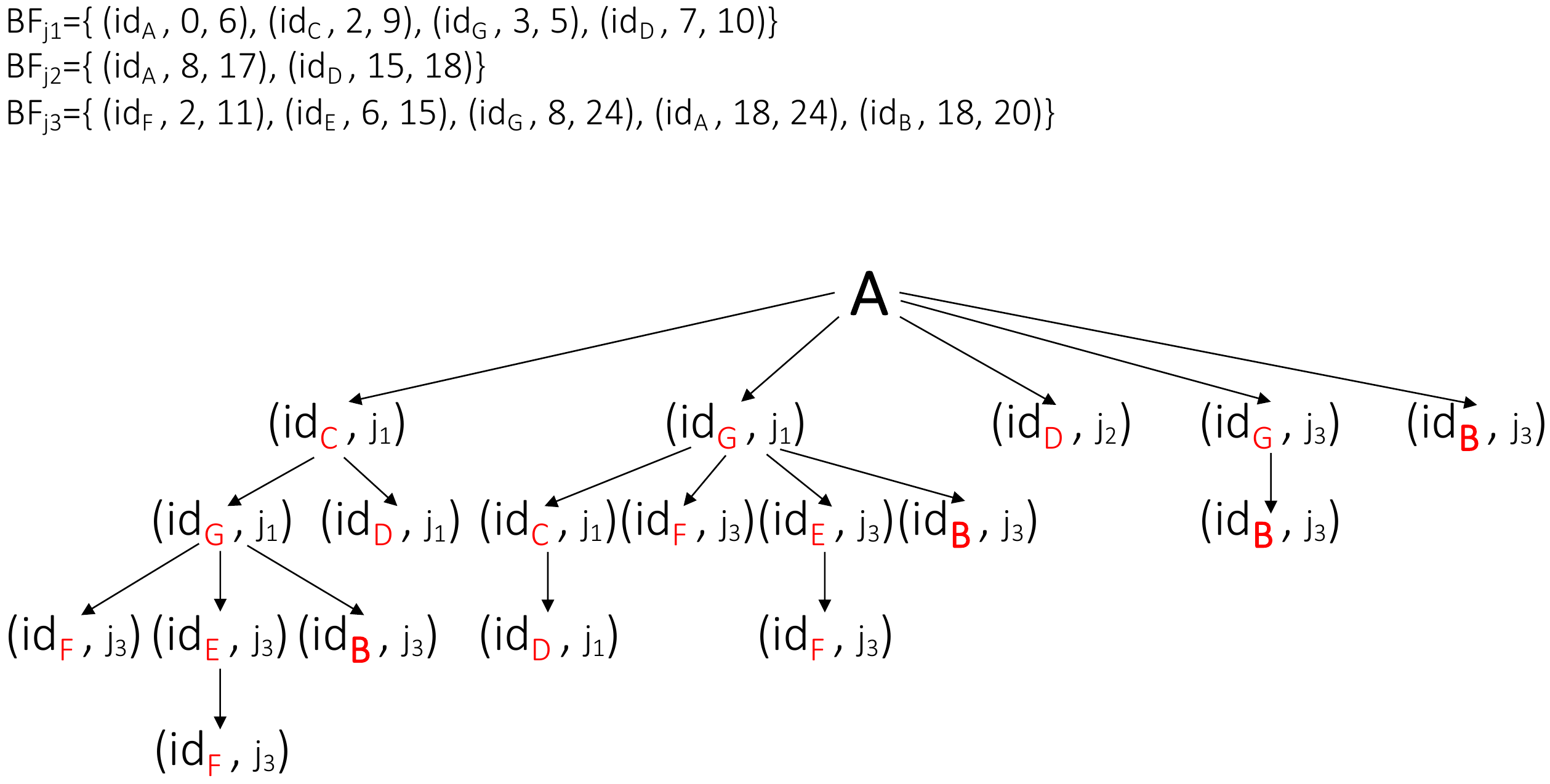}
\end{center}
\caption{Example of connection logfiles from three base stations and the respective proximity tree obtained from $prox\_tree(A, A, B, 0)$. It outcomes three different proximity chains $\langle A, C, G, B \rangle$, $\langle A, G, B \rangle$ and $\langle A, B \rangle$.}
\label{fig:tree}
\end{figure}

In Figure \ref{fig:tree} we give a toy example of our recursive algorithm with seven users $A, B, C, D, E, F, G$, three base stations $BS_{j_1},BS_{j_2},BS_{j_3}$ and times as integers in $[0;24]$. We show the content of connection logfiles from the three base stations and the proximity tree from user $A$ to user $B$ that has been generated by computing $prox\_tree(A, A, B, 0)$. We observe in Figure \ref{fig:tree} that two users might be in contact around different base stations. Indeed, the resulting proximity chains are $\langle A, C, G, B \rangle$, $\langle A, G, B \rangle$ (with users A and G in proximity around $BS_{j_1}$), $\langle A, G, B \rangle$ (with users A and G in proximity around $BS_{j_3}$) and $\langle A, B \rangle$. In case there are evaluated as infection chains, users $C$ and $G$ might also be infected.

\paragraph{Algorithm optimization}

With respect to performance, one could consider computing the algorithm on the opposite way, namely with input $B$ as root. To do so, the algorithm should be modified so that time is considered backwards. It starts at ending time (24 for our toy example) and we build the proximity tree by going back in time. We consider as $reverse\_prox\_tree$ this reverse recursive algorithm.

\begin{figure}[h]
\begin{center}
\includegraphics[width=0.4\textwidth]{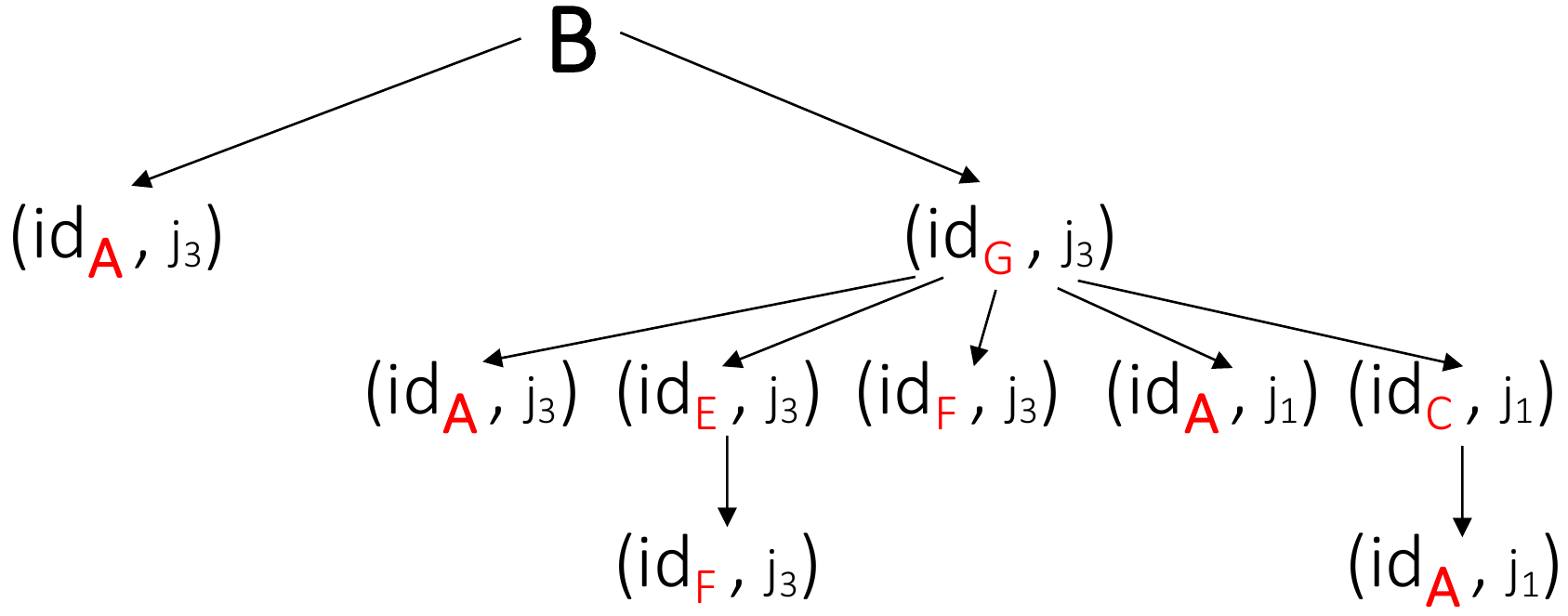}
\end{center}
\caption{Example of a proximity tree obtained from $reverse\_prox\_tree(B, A, B, 24)$ with the same toy example than Figure \ref{fig:tree}. It generates three different proximity chains $\langle A, C, G, B \rangle$, $\langle A, G, B \rangle$ and $\langle A, B \rangle$.}
\label{fig:tree2}
\end{figure}

In Figure \ref{fig:tree2} we show the proximity tree obtained after computing $reverse\_prox\_tree(B, A, B, 24)$ from user $B$ considering the time backwards. As expected, the resulting proximity chains are the same than in Figure \ref{fig:tree} but we remark that the resulting tree is smaller than the one obtained in Figure \ref{fig:tree}. In this specific toy example we notice that obtaining the proximity tree was made faster by reversing our algorithm.

Another aspect we could consider while comparing the two resulting trees, is that the order the tree is being build and the proximity chain obtained is also reversed. Indeed, in Figure \ref{fig:tree} we obtain first $\langle A, C, G, B \rangle$ then $\langle A, G, B \rangle$ (via $j_1$), $\langle A, G, B \rangle$ (via $j_3$) and finally $\langle A, B \rangle$. In Figure \ref{fig:tree2} we see that we obtain the chains in the exact opposite order with $reverse\_prox\_tree$.
Still aiming to optimize the computation time of our algorithm, in particular when dealing with large numbers of users and base stations, one could simultaneously start the tree generation using the algorithm and its reversed version. For both cases the tree propagates and every time we find a proximity chain in the tree (meaning $N=B$ or $N=A$ for $reverse\_prox\_tree$) we could store the chain in a set $S$ (or $S^\prime$ for $reverse\_prox\_tree$).
Then for each round (i.e \textit{for} iteration) we test if the two sets have a common element. If not, we continue. In case they have a common proximity chain, we could stop both algorithms and the complete set of proximity chains from users $A$ to $B$ is composed of the addition of sets $S$ and $S^\prime$.

To illustrate the approach of computing both versions at the same time and, as argued, gain on performance, one could explain:
\begin{itemize}
\item if you throw one stone into the water and you want the resulting waves to reach a point in $r$ meters distance, then the circle at the end will encompass many square meters.
\item if you throw two stones into the water (one at the original position, the other one at the position you want to reach), the intersection of the resulting waves propagation will be approx. at a distance $r/2$ meters.
\item adding the area of these two circles shall be much smaller than the circle's area obtained with one stone.
\end{itemize}
For example, with $A=\pi \times r^2$ and $r=10$ $A=314.159$, and with $r=5$ the area of the two circles is altogether approximately 160!

%- if you throw one stone into the water and you want to have the resulting waves to reach a point in $r$ meters distance, then the circle at the end will have rather much square meters.\\
%- if you throw two stones into the water (one at the original position, the other one at the position you want to reach), the intersection of the resulting waves propagation will be approx. at a distance $r/2$ meters.\\
%- Adding the area of these two circles should be much smaller than of the circle's area with one. \\
%For example, with $A=\pi \cdot r^2$ and $r=10$ $A=314.159$, and with $r=5$ the area of the two circles is altogether approx. 160!\\

Another level of optimization could be considered in order to identify some of the proximity chains faster as for instance to support the start of a localized quarantine immediately.
Instead of storing the chains into $S$ and $S^\prime$, at each propagation round we look at the chains while they are processed so that we stop both algorithms when:
\begin{itemize}
\item $prox\_tree$ has built a path $\langle A,X_1,\dots,X_i\rangle$
\item $reverse\_prox\_tree$ has built a path $\langle B,X_n,\dots,X_j\rangle$
\item and it holds $X_i==X_j$
\end{itemize}
Then the two parts of the proximity chain could be concatenated to create the proximity chain $\langle A,X_1,\dots,X_i==X_j,\dots,X_n,B\rangle$

We could refer to Table \ref{tab:opt} to see that if we perform both algorithms at the same time in the toy example configuration, we could retrieve the proximity chain $\langle A,C,G,B\rangle$ faster with this second level of optimization.

In Table \ref{tab:opt} we could observe in detail how we retrieve the proximity chains using the two versions of Algorithm \ref{algo:F} and the optimization with the toy example's configuration. As stated previously, $reverse\_prox\_tree(B,A,B,24)$ was executed way faster than $prox\_tree(A,A,B,0)$. Indeed, the original algorithm ended after 18 rounds while the reverse one stopped after the $9^{th}$ round. Since it is not possible to predict which of the two will finish processing first, computing both in parallel will optimize the retrieving. As for the second level of optimization, concatenating two parts of proximity chains allows to retrieve $\langle A,C,G,B\rangle$ at round 2 while discovered at round 6 with $prox\_tree$ and round 9 with $reverse\_prox\_tree$. It is of value especially when proximity chains are composed by a high number of intermediate users.

\begin{table*}[h]
%\caption{Running times of the two set relations in seconds with $\nkey$ selected in $[5.10^2 ; 2.10^3]$.}
\caption{Construction of the proximity tree round by round with $prox\_tree$ and $reverse\_prox\_tree$ and how the optimization could be applied.} 
\centering
%\begin{adjustbox}{width=0.7\textwidth}
\begin{tabular}{cccl}
\toprule
	Round&	$prox\_tree$ &	$reverse\_prox\_tree$ &	With optimization\\
	\midrule
	1	&	$C$&$A, \mathbf{\langle A,B\rangle}$&	 $\mathbf{\langle A,B\rangle}$ from $reverse\_prox\_tree$.\\
	&&&from $reverse\_prox\_tree$.\\
	\midrule
	2	&	$G$&$G$&	$\mathbf{\langle A,C,G,B\rangle}$ from concatenation of $\langle A,C,G\rangle$ from\\
		&&&$prox\_tree$ and $\langle G,B\rangle$ from $reverse\_prox\_tree$.\\
		\midrule
	3	&$F$&$A,\mathbf{ \langle A,G,B\rangle}$&$\mathbf{\langle A,G,B\rangle}$ from $reverse\_prox\_tree$.\\
		\midrule
	4	&	$E$&$E$&\\
		\midrule
5	&	$F$&$F$&\\
\midrule
6	&	$B, \mathbf{\langle A,C,G,B\rangle}$&$F$&\\
\midrule
7	&	$D$&$A, \mathbf{\langle A,G,B\rangle}$&$\mathbf{\langle A,G,B\rangle}$ from $reverse\_prox\_tree$.\\
\midrule
8	&	$G$&$C$&\\
\midrule
9	&	$C$&$A, \mathbf{\langle A,C,G,B\rangle}$&\\
\midrule
$\dots$	&	$\dots$&$\dots$&\\
\midrule
14	&	$B, \mathbf{\langle A,G,B\rangle}$&-&\\
\midrule
$\dots$	&	$\dots$&$\dots$&\\
\midrule
17	&	$B, \mathbf{\langle A,G,B\rangle}$&-&\\
\midrule
18	&	$B,\mathbf{ \langle A,B\rangle}$&-&\\
\midrule
\bottomrule 
\end{tabular}
%\end{adjustbox}
\label{tab:opt}
\end{table*}

The performance gain obtained with our two levels of optimization is downplayed due to the extreme smallness of logfiles in our toy example. 
One could easily imagine that applied to real life scenario and 	big data these optimizations are highly performance saving. For example, in another scenario dealing with mobile connection logfiles \cite{DBLP:conf/cecc/TajanW19}, authors propose to process on these logfiles and therefore Bloom filters up to $10^6$ elements.

\paragraph{Algorithm decentralization}
The European PEPP-PT consortium is advocating a decentralized approach as well as the DP3T protocol \cite{troncoso2020decentralized} which relies on Bluetooth, and also as \cite{cryptoeprint:2020:531} where decentralization has been investigated. With our presented optimization, we could integrate such construction by introducing two additional parties besides the ones already presented. We precise that these two additional parties should be extremely powerful in terms of computation and perform parallel computing such as server farms or clusters:
\begin{itemize}
\item \textit{Computing party 1} which runs $prox\_tree$
\item \textit{Computing party 2} which runs $reverse\_prox\_tree$
\end{itemize}
This way the agency is only receiving per round the values for $X_i$ (from \textit{computing party 1}) and $X_j$ (from \textit{computing party 2}) and comparing if $X_i==X_j$. Only in the case $X_i==X_j$ we obtain that \textit{computing party 1} is sending $\langle A,X_1,\dots, X_i\rangle$ and \textit{computing party 2} sending $\langle X_j,\dots, X_n, B \rangle$ to the agency. With such a construction, multiple parties are involved in the computation and the whole effort does not rely on the government agency.

\paragraph{Algorithm complexity}
One could easily see by analyzing the obtained results in Figures \ref{fig:tree} and \ref{fig:tree2} that the size of the resulting tree will depend on the size of the base stations' logfiles. These logfiles will naturally depend on the amount of users and thus connections during the particular time. The more base stations and users there are, the more logfiles will be numerous and fully filled. In our toy example, we have 11 connection entries in all combined base stations as displayed in Figure \ref{fig:tree}. They result in a tree with respectively 19 and 10 nodes by computing $prox\_tree$ and $reverse\_prox\_tree$. We also recall that in case we find the final user of the wanted infection chain (user $B$ in our example) in the tree, the algorithm reaches a break instruction and therefore the respective sub-tree is no longer explored. A high activity of this particular user could then reduce the tree's spreading. As seen previously, one of the two algorithms will be faster to execute without being able to predict which one and applying the presented optimization could reduce the complexity to the faster one.

\subsection{Proximity Chain Evaluation}
From all the proximity chains $\langle A, id_{X_1}, \dots, id_{X_n}, B\rangle$ obtained by performing the aforementioned protocol, the government agency should determine if users $X_i$ might also be infected. To do so, the agency could estimate the users' probability of being infected and compare it to a threshold (i.e. $Pr(X_i)>Tr$). Such a probability obviously depends, among others, on the respective neighbors within the chain. We consider the probability value computed as a function $infection(previous\_node, contact\_time, contact\_distance,\\ reproduction\_number,$ $saturation)$ where \textit{saturation} shall denote the percentage of infected persons within the human population of a region, which obviously changes over time. 

More precisely, in Germany the reproduction number $R$, which is defined as the mean number of people infected by a case, was 3 at the beginning of the COVID-19 crisis and by 17.04.2020 could be reduced to $0.7$ (and meanwhile $R=1.1$). Clearly this number is only an average but still indicates that inference from a proximity chain to an infection chain very much depends on the concrete time and location entities met during the pandemic wave. Similar numbers also exist for other countries as for instance $R= 0.8$ for Belgium at 17.04.2020. Another important observation is that since a proximity chain can easily build up over a period of weeks, $Pr(X_i)$ may significantly vary. But only if all probabilities are larger than $Tr$ the agency can at least argue having identified a possible infection chain.

%After having identify all the possible paths between users A and B, we could use an ``infection equation" to evaluate the plausibility of them to have really happened.

It goes without saying that it is out of scope to determine the $infection$ function. On the one hand, specialists emphasize the high contagiousness of the virus but on the other hand, having two users connecting to the same base station at the same time does not necessarily imply any physical contact between the two.

%There exists several approaches that could be of interest.
%In \cite{van2017distributed}, the authors propose the following formula based on the gossip theory to represent the virus spreading:
%\begin{equation}\label{eq:s}
%s = e^{-(1/p_{stop}+1)(1-s)}
%\end{equation}
%computing $s$ the fraction of nodes that will remain ignorant to infection and $p_{stop}$ the probability for a node to stop spreading the virus.

%We could also draw from another approach by Westhoff $\etal$ \cite{DBLP:journals/sigmobile/LamparterPW03} and apply the following probability equation:
%\begin{equation}
%Pr(E^1_p \ \land \dots \land \ E^n_p) = e^n
%\end{equation}
%to estimate if user $A$ has indeed been infected by user $B$ (here $E^1_p$ can be modeled with the Gossip probability from \cite{van2017distributed}).

Without being able to determine the exact probability of a user to be infected by another one, we could propose a model to evaluate the probability of a proximity chain becoming an infection chain.
First, we know that users $A$ and $B$ are infected and we would like to determine if user $B$ has been infected due to user $A$ or via another chain and other infection events. Therefore, applying probability theory to such a problem is relevant and reflects the \textit{chain} characteristic of it.

We define as $Pr(X_i)$ the following conditional probability\\ $P(X_i | X_{i-1})$ of the event "$X_{i-1}$ has infected $X_i$ knowing that $X_{i-1}$ is already infected".
%Let it be two independent events:
%\begin{itemize}
%\item[] \textit{event 1}: $X_{i-1}$is infected and has infected $X_i$
%\item[] \textit{event 2}: $X_{i}$ is infected and has infected $X_{i+1}$
%\end{itemize}
It holds that $Pr(X_1\cap \dots \cap \ X_n) = \displaystyle \prod_{i=1}^{n} Pr(X_i)$. Considering a proximity chain $\langle A, X_1, \dots, X_n, B\rangle$, there is a clear tendency that the overall probability to have user $B$ infected due to user $A$ is inversely proportional to the length of the proximity chain.
We propose the following probability model for evaluating a proximity chain:
%\begin{theorem}
%For each  $\langle A,X_1,\dots,X_n,B\rangle$ if $T_r \leq Pr(X_1) . \cdots . Pr(X_n)$ then $[A,X_1,\cdots,X_n,B]$
%\end{theorem}
%\begin{description}
%\item For each  $\langle A,X_1,\dots,X_n,B\rangle$, if $\displaystyle \forall i \in [1;n]  Pr(X_i) \geq T_r $\\ then $[A,X_1,\dots,X_n,B]$.
%\end{description}
\begin{multline}\label{eq:eval}
\text{For each } \langle A,X_1,\dots,X_n,B\rangle \text{, if }\displaystyle \forall i \in [1;n]  Pr(X_i) \geq T_r \\ \text{ then }[A,X_1,\dots,X_n,B]
\end{multline}
%$$\text{For each }  \langle A,X_1,\dots,X_n,B\rangle \text{ if } T_r \leq Pr(X_1) . \cdots . Pr(X_n) \text{ then }[A,X_1,\cdots,X_n,B]$$

%Even without being able to estimate the probability of infection between two users connecting to the same base station at the same time, 
%Even without any epidemiology expertise one could notice that more an infection chain is short, more it is probable to happen.

The proximity tree obtained at the previous stage of the protocol contains nodes with users' credentials and only these usernames are revealed. It is only in case a proximity chain turns out to be an infection chain, that the agency will request from the telco company the real identities of the users composing the chain. 
Therefore, users' identity are solely revealed in case of \textit{infection} function outcomes so. Moreover, we recall that during the overall process no additional location information of other users listed in the mobile operator's logfile are revealed to the agency.

%Then, with a threshold $Tr < Pr(E^1_p \ AND \dots AND \ E^n_p)$, reveal the chain $\{A, X_1, \dots, X_m, B\}$ and contact each intermediate user.
Another way to tune $prox\_tree$ and make the overall computation more salable could be, during the computation of $prox\_tree$ and $reverse\_prox\_tree$, to only consider such paths in the proximity tree as long as they still fit the criterion to also be an infection chain. It could consists of having the testing from equation \eqref{eq:eval} at line \ref{line20:algo} from Algorithm \ref{algo:F} and a \textit{break} instruction in case the test is not fulfilled.

\subsection{Recursivity of the Infection Detection}
\begin{figure}[h]
\begin{center}
\includegraphics[width=0.35\textwidth]{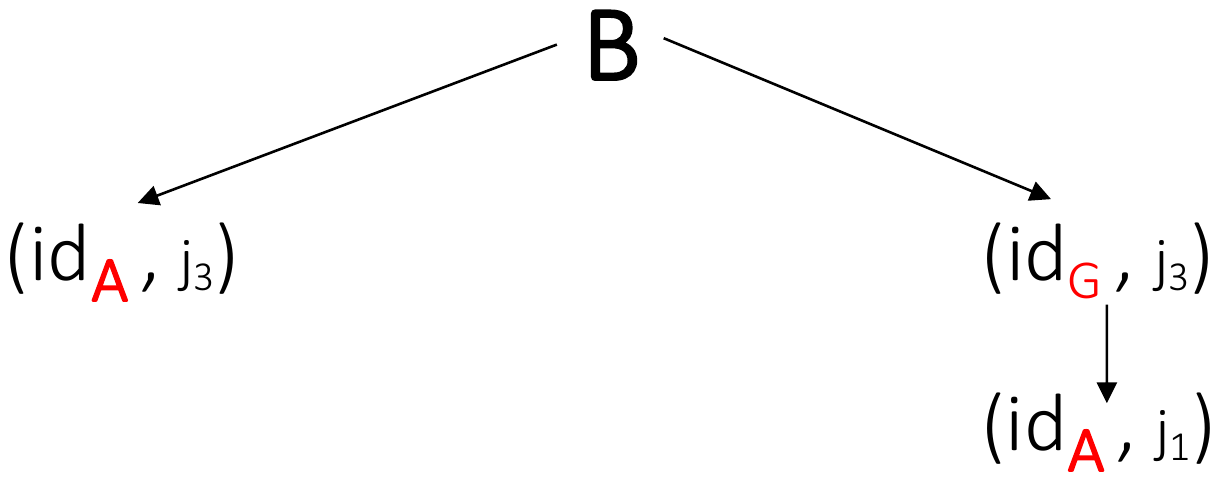}
\end{center}
\caption{Example of a proximity tree from user $B$ to user $A$ obtained from $prox\_tree(B, B, A, 0)$. It results in two different proximity chains $\langle B, A \rangle$ and $\langle B, G, A \rangle$.}
\label{fig:tree3}
\end{figure}

One may notice that a trivial optimization would be to switch users $A$ and $B$ in the sense that ``infection of user $A$ is coming from user $B$". In Figure \ref{fig:tree3} we show the proximity tree obtained from our algorithm by computing $prox\_tree(B, B, A, 0)$ with our toy example logfiles. We notice that it results in a very different tree than in Figure \ref{fig:tree} obtained by $prox\_tree(A, A, B, 0)$. In case the government agency holds some information on the infection time of users $\A$ and $\B$, for example that user $\A$ has been infected before user $\B$, only one direction should be considered by the agency.

To be the most efficient, the government agency should perform a final step in the protocol. All the users identified as infected at the previous stage (i.e. all $X_i$ where $Pr(X_i) > Tr$) should be considered as new users $A$ and respectively $B$ in the proposed solution. Indeed, our protocol is initiated with users tuples $(A,B)$ already identified as infected by the agency. The freshly identified users are thus incrementing the list of known infected persons and the protocol should be applied to them to optimize the search. In such a way, the most infected users could be identified and contacted.

\subsection{Discussion on Location Privacy}
We argue that the proposed solution provides privacy for the users by three different means. Firstly by using only personal credentials as usernames and secondly thanks to the Bloom filter's construction and its obfuscation feature.
Indeed, as explained previously, the real identities of users are not provided and stored in the Bloom filters nor the logfiles. The telco company uses usernames to distinguish users and the private mapping will be provided to the government agency solely on-demand, when a user is identified as being part of an infection chain.

The second aspect of location privacy is given by the Bloom filters based approach from \cite{DBLP:conf/icete/TajanWA20} which allows to compute relations among logfiles while keeping these data sets private. We recall that such an approach uses an HMAC function instead of a bunch of public hash functions and therefore only the telco company could create the Bloom filters and no other party. To this extent, the government agency could not try to retrieve locations of a specific user by generating a Bloom filter with a unique element and performs the inclusiveness relation between this Bloom filter and the ones from base stations. For that reason, using secret keys to generate a valid Bloom filter enhances the privacy aspect of the protocol. Finally we recall that secret keys are generated and stored only at the telco company side and are not required by the government agency to perform our protocol.

The third aspect of location privacy consists of having no other party than the provider itself (which anyhow has this information) gets the location data of the users. This can be easily done by not revealing which $BF_i$ comes from which $BS_i$. This way, the only information revealed to the authority is the contact information of users having entered the same cell during the same time interval. Providing the concrete location information of this cell is totally irrelevant for the authority to compute the proximity resp. infection chain.

%% file: conclusion.tex
\section{Conclusion}
We proposed in this work to use the Bloom filter approach from \cite{DBLP:conf/icete/TajanWA20} for a real life use case, similarly to \cite{DBLP:conf/cecc/TajanW19} where we applied it to a post-mortem mobile device tracking scenario. Our detailed protocol supports a government agency to track possible COVID-19 infection chains and therefore identify plausible infected mobile users. Throughout the entire protocol, the agency will only handle usernames which do not allow to retrieve the users' identities and therefore their privacy will be preserved. Solely in the case of possible infection by the life-threatening SARS-CoV-2 virus, real identities will be revealed to the agency, that will be able to contact them and provide medical support. In such way, the telco companies act GDPR compliant and could still guarantee a certain level of location privacy to their clients.
We could stress that if data stem from the `in proximity' mobile telco's logfile, it means that two devices have been in the same transmission range of a base station. In the worst case they can still have a $2\times r$ distance (easily 500 m or more). However, if the same approach can be applied to the RSSI based Swarm-mapping approach for Android or iOS collected data then `in proximity' has a much better accuracy \cite{DBLP:journals/insk/DheinG17}. In particular also the WiFiLocationHarvest file of each mobile device contains timestamp, latitude, longitude, trip-id, speed, course at an amazing accuracy which comes close to the accuracy required to check if two devices got nearer than 2 m (infection distance). And, moreover, compared to the promoted App based approach with Bluetooth from \textit{Germany Fraunhofer Institutes} and others in the RSSI based approach the mobile's WLAN and Bluetooth can be off, and yet, simply due to the measured RSSI from the access point the approach provides the location data of the devices equipped with such modern mobile operating systems.

To conclude, our approach may be a good starting point for debating a reasonable GDPR compliant detection of COVID-19 infection chains since we argue it does not provide additional privacy-leakage to other parties than those who already have the knowledge of our location data. 